\documentclass[runningheads]{llncs}
\usepackage[T1]{fontenc}
\usepackage{silence} 
\usepackage{amsmath} % equations
\usepackage{amssymb} % Real numbers R
\usepackage{graphicx}
\usepackage{caption}
\usepackage{subcaption}
\usepackage{tikz}
\usetikzlibrary{shapes.geometric, arrows.meta, positioning, calc, positioning, backgrounds}
\usepackage[hyphens]{url}
\usepackage{hyperref}

\usepackage{algorithm} %pseudocode
\usepackage{algpseudocode}
\usepackage{placeins}
\usepackage{makecell}

\usepackage{color}

\begin{document}
\title{A Graph-Based Approach to Alert Contextualisation in Security Operations Centres}
% \title{A graph-based approach to correlate alerts with known incidents in a security operation centre}
%\title{Alert triage in a SOC using alert graphs}
%
\titlerunning{Graph-Based Alert Contextualisation}
\authorrunning{Eckhoff et al.}
% If the paper title is too long for the running head, you can set
% an abbreviated paper title here
%

\author{Magnus Wiik Eckhoff\inst{1,2}\and
Peter Marius Flydal\inst{3}\and 
Siem Peters\inst{3}\and 
Martin Eian\inst{3}\and
Jonas Halvorsen\inst{1} \and
Vasileios Mavroeidis\inst{2} \and
Gudmund Grov\inst{1,2}
}

%
% \authorrunning{M. Eckhoff et al.}
% First names are abbreviated in the running head.
% If there are more than two authors, 'et al.' is used.
%
\institute{Norwegian Defence Research Establishment \and
University of Oslo \and mnemonic AS}
\maketitle              % typeset the header of the contribution
\begin{abstract}
% Option 1
Interpreting the massive volume of security alerts is a significant challenge in Security Operations Centres (SOCs). Effective contextualisation is important, enabling quick distinction between genuine threats and benign activity to prioritise what needs further analysis. 
This paper proposes a graph-based approach to enhance alert contextualisation in a SOC by aggregating alerts into graph-based alert groups, where nodes represent alerts and edges denote relationships within defined time-windows. By grouping related alerts, we enable analysis at a higher abstraction level, capturing attack steps more effectively than individual alerts. Furthermore, to show that our format is well suited for downstream machine learning methods, we employ Graph Matching Networks (GMNs) to correlate incoming alert groups with historical incidents, providing analysts with additional insights.


\end{abstract}
\section{Introduction}\label{sec:intro}

Modern cyber attacks come with substantial financial and operational costs, and it is generally accepted that purely preventive measures are not sufficient to stop them~\cite{vielberth2020security,fleck2024cybercrime}. 
% alternativ 1
This requires capabilities for detecting and responding to incidents \cite{nelson2025incident}, an activity that is usually carried out within Security Operations Centres (SOCs). 
% alternativ 2
% Therefore, the ability to detect and respond to incidents is needed \cite{nelson2025incident}, an activity that is typically performed by Security Operations Centres (SOCs). 

% Why this is hard
To respond to suspicious behaviour that has been detected, one must first analyse and understand what is likely to have happened. This is a challenging and labour-intensive task for several reasons.
First, security alerts must be quickly assessed and prioritised to identify what needs further analysis~\cite{lin2018retrieval}. 
This process typically involves examining alerts, correlating them with relevant sources, and filtering out unrelated information. 
The sheer number of alerts -- most of them false -- coupled with limited available analyst time, has led to a problem known as \emph{alert fatigue}\footnote{Alert fatigue describes how security analysts become both overwhelmed and desensitised, which may result in true positive alerts being missed, efficiency reduced, and the analyst suffering burnout.} in many SOCs \cite{tariq2025alert,zhong2016automate}. 

% Part of larger + alert triage 
Any support and automation that can be provided to help analyse and understand the alerts could have a massive impact on a SOC. In this paper, we address the initial phase of such analysis
%, often called \emph{alerttriage}~\cite{lin2018retrieval},
in two ways. 
First, we address the problem of alert correlation, meaning the ability of an analyst to examine a group of alerts at the same time instead of handling each alert individually. 
Second, we propose a method to match graph-based alert groups with historical ones, enabling insights from past events to serve as the starting point for further analysis.

% Observations
Our work is based on a set of (informal) observations made from how SOCs operate in practice.
Firstly, whilst an alert in isolation is often connected to a (low-level) indicator of compromise (IoC)\footnote{Examples of IoCs are IP addresses, hostnames or even known vulnerabilities.}, a combination of alerts can indicate more abstract behavioural patterns\footnote{A single alert may also be an indicator of behaviour, but even so, combined with other alerts will still provide a more abstract view.}. It is known that IoCs are easy for an attacker to change, and to detect attacks from advanced threat actors, IoCs are not a suitable abstraction level to work on when trying to find similarities across time. Behavioural patterns, on the other hand, are much harder to change, and therefore more likely to remain the same~\cite{bianco2013pyramid}. 
Secondly, while we are unlikely to have much knowledge of the new alerts, we are likely to have more information about previous incidents and campaigns. This could be anything from what phase of the attack an alert group is likely to belong to, to detailed Cyber Threat Intelligence (CTI) -- for instance, what threat actor we are dealing with, or what the goal of the attack might be.

% RQ 1,2 structure in alerts
By connecting alerts in a structured way, we hypothesise that we can capture high-level behavioural patterns that can be matched with, and used to exploit patterns of known previous incidents. We argue that graphs are a natural representation as they can efficiently represent key connections between the correlated alerts. Moreover, an attack happens in different phases, and alerts can be correlated at different levels: they can be correlated to relate different phases of an attack or correlate alerts within the same phase. Our focus is on the latter, which leads us to our first two research questions: 
\begin{description}
    \item[\textbf{RQ1:}] How naturally can related alerts be combined into graph-based alert groups such that each alert in a group belongs to the same phase of an attack?
\end{description}

\begin{description}
    \item[\textbf{RQ2:}] How to automatically combine alerts from the same attack step together while minimising the number of unrelated alerts in the group?
\end{description}
% RQ 3 noise 
We note that threat detection in cybersecurity has to deal with a considerable amount of noise, meaning that any grouping method will likely result in groups containing false positives and genuine alerts. Consequently, a method that attempts to match such a graph-based alert group to another one representing historical incidents needs a level of noise resistance. One promising approach that has been applied to related problems in other domains is \emph{Graph Matching Networks} (GMNs) ~\cite{li2019graph}, which we explore in our final research question:
\begin{description}
    \item[\textbf{RQ3:}] To which degree can Graph Matching Networks correlate current alert groups to related historical incidents?
\end{description}
% contributions of this paper
The contributions of this paper follow the research questions and are two-fold: 
(1) we provide a formal account of alert graphs and describe a method for creating graph-based alert groups from a stream of alerts; (2) we apply and adapt the graph-based machine learning method \emph{Graph Matching Network} to our problem domain, and show that it is able to match related graph-based alert groups in order to correlate new alert groups with known incidents, thus providing a promising approach to support security analysts and reduce alert fatigue.

% organized as follows.
The paper is organised as follows: 
section~\ref{sec:background} provides a brief overview of the background;
section~\ref{sec:alert_structure} gives a formal account of alert graphs and a method for generating graph-based alert groups;
section~\ref{sec:GMN} outlines our approach in using Graph Matching Network to correlate current and historical graph-based alert groups;
section \ref{sec:evaluation} contains the evaluation of our approach, followed by a discussion in section \ref{sec:discussion} and a conclusion of the paper in section~\ref{sec:conclusion}.

\section{Background}
\label{sec:background}
The problem of alert fatigue in SOCs is well known (see e.g.~\cite{tariq2025alert}),
where a large proportion of the alerts are either false or not important~\cite{alahmadi202299}. In one study, it was shown that $62\%$ of all alerts in SOCs are  ignored~\cite{davis2024MSSP}. Since alert fatigue has been a persistent and longstanding issue, different solutions and methods have been developed to mitigate its impact.
% and over 56\% of large organizations are dealing with more than 1,000 security alerts on a daily basis \href{https://www.darkreading.com/cyber-risk/56-of-large-companies-handle-1-000-security-alerts-each-day}{link}, making effective alert triage a major operational challenge.

% Moreover, over 56\% of large organizations are dealing with more than 1,000 security alerts on a daily basis \href{https://www.darkreading.com/cyber-risk/56-of-large-companies-handle-1-000-security-alerts-each-day}{link}, making effective alert triage a major operational challenge. The time burden associated with these alerts is also considerable, with most respondents in another survey indicating that investigating a single alert takes more than 10 minutes \href{https://www.criticalstart.com/wp-content/uploads/CS_MDR_Survey_Report.pdf}{link}. In 38\% of the SOCs surveyed, the high volumes of alerts combined with limited analyst resources have led organizations to hire additional analysts or deliberately ignore specific high volume alert types altogether.

% % Filtering alerts
% Since alert fatigue has been a persistent and longstanding issue, different solutions and methods have been developed to mitigate its impact. An option to combat alert fatigue is to classify alerts by filtering out clearly benign alerts before sending them to the SOC. Closely related to alert classification is alert prioritization. The idea is to rank alerts based on predicted suspiciousness, to push more suspicious alerts to the SOC first, leveraging the benefits of both machine learning and humans \href{https://dl.acm.org/doi/pdf/10.1145/3695462}{link}. 

% Automated alert filtering and priorization 
One approach has been to filter out unrelated alerts or prioritise important alerts. An example is the \emph{Automated Alert Classification and Triage} (AACT) system, which leverages machine learning to replicate SOC analysts’ triage decisions~\cite{turcotte2025automated}. The system enables real-time classification and prioritisation of alerts, reducing analyst workload by automatically dismissing clearly benign alerts and prioritising critical ones. 
% The method Carbon Filter~\cite{oliver2024carbon}, efficiently identifies clear false positive alerts by analysing the context in which the process was initiated, such as the command-line arguments used to start it. The strong performance of Carbon Filter significantly improved the signal-to-noise ratio of alerts without negatively affecting triage performance. 
\emph{That Escalated Quickly} (TEQ)~\cite{gelman2023escalated} is a machine learning framework that predicts alert- and incident-level actionability, 
%reducing response time to actionable incidents, suppressing false positives while maintaining a high detection rate, and decreasing the number of alerts analysts must investigate per incident. 
while \emph{RAPID}~\cite{liu2022rapid} is a collaborative real-time alert investigation system that attempts to reduce redundant triage work and dynamically prioritise investigations. 
%allowing efficient provenance analysis without relying on predefined threat ontologies.
% Another approach combines anomaly scoring over provenance graphs with high-level alert causality %reducing false positives, speeding up investigations, and minimising log retention while preserving multi-stage attack traceability
% ~\cite{hassan2019nodoze,Hassan2020Tactical}.
% Why we differ
These techniques focus on filtering out uninteresting alerts or prioritising interesting alerts before any further analysis. In contrast, our approach contextualises all alerts, enabling relevance to be determined based on this context rather than upfront filtering.   

% % prioritizing resources in a soc ✂️
% Another way to combat alert fatigue is by improving the resource allocation of a SOC. \href{https://ink.library.smu.edu.sg/sis_research/4666/}{link} aim to optimize the assignment of intrusion alerts to a limited number of cyber analysts. The authors propose a Cyber Screening Game (CSG) model, an NP-hardness proof to compute the optimal defender strategy, and develop optimal and scalable heuristic algorithms to allocate analyst resources effectively under adversarial conditions.

% alert grouping
Another way to address alert fatigue is to combine related alerts into a single group. As a result, an analyst can work with a group of alerts instead of the alerts individually.
% Grouped alerts can reduce the required amount of manual investigation analysis or be used for classification and triage, as has been done on single alerts.
It has been argued that representing alerts as graphs can improve the workflow of analysts, improve alert prioritisation and classification, reduce analyst workload, and detect threats that would have gone unnoticed if investigated in a single alert~\cite{jalalvand2024alert}.
Graphs and graph-based learning have been studied to correlate different stages of an attack
\cite{liu2022behind}. Here, a hyperconnected graph is generated with nodes representing alerts and entities (e.g., IP addresses, accounts) and edges representing the relationships between the nodes. Known attack patterns are stored in a knowledge pool, and \emph{K-Nearest Neighbours}
(KNN) uses these patterns to prioritise alerts. A comparable approach \cite{fredj2015realistic} represents alerts as nodes and transitions between them as edges, and uses \emph{Markov chains}~\cite{norris1998markov} to detect attack chains. There are also graph-based commercial solutions \cite{palo,google_alert_grouping,sentinel} to discover such multi-stage attack chains.
%One approach to graph-based alert correlation constructs graphs with alerts as nodes and the transition between different types of alerts as edges~\cite{fredj2015realistic}. From this, attack scenarios are detected using Markov chains~\cite{norris1998markov}
%\textcolor{red}{One approach to graph-based alert correlation constructs graphs with alerts as nodes and the transition between different types of alerts as edges~\cite{fredj2015realistic}. From this, attack scenarios are detected using Markov chains~\cite{norris1998markov}.
%identifies matches within the graph to detect potential attacks. K-Nearest Neighbours (KNN) is then applied to compute kill-chain reachability, assigning a relevance score to prioritise alerts.
%\textcolor{red}{
%Other private actors also provide services for grouping or linking alerts and %logs~\cite{palo,google_alert_grouping,sentinel}.
%These works share the most similarities with this paper, but differ in abstraction level. The %approaches aim to find entire attack chains within a given subgraph and are dependent on doing %so. Meanwhile, our approach focusses on discovering individual attack steps in isolation.  }
Whilst sharing the hypothesis that graphs are a suitable representation for alerts \cite{jalalvand2024alert}, our work deviates from the above graph-based approaches by focusing on correlating alerts within a single phase of an attack.

% Normalising alerts from different sources, by predicting which attack step any given alert is, can improve the ability to group related alerts together~\cite{wei2023fusing}. 

One prominent non-graph-based approach for alert correlation is based on time, where alerts are grouped based on a set time interval~\cite{landauer2022dealing}.
This method does not require defining attack scenarios or specific alert formats, and has been used to extract higher-level alert patterns. However, the approach preserves no structure between alerts and has known limitations with simultaneous attacks.

\section{Introducing and generating alert graphs}
\label{sec:alert_structure}

To reduce alert fatigue in a SOC, it can be useful to divide incoming alerts into the following three categories:
\begin{enumerate}
    \item alerts that contain enough information to be actionable;
    \item alerts that never contribute to analysis;
    \item alerts that may be part of a broader analysis that leads to something actionable.
\end{enumerate}
Alerts of the first type should be dealt with automatically using \emph{SOAR} (Security Orchestration, Automation and Response) or similar methods, and are not part of the issue we study in this paper. Alerts of the second type are essentially noise and should be tuned out of the alert pipeline. These two categories contain the alerts that are always either true positives or true negatives\footnote{Note that both of these two categories are highly non-trivial to deal with in practice, but they are not within the scope of this paper.}.

Alerts of the third kind are our focus, as they are the ones for which false positives and false negatives necessarily exist, because no amount of alert-level tuning or automation has even a theoretical chance of handling them -- they are, by our definition, ambiguous in isolation. To identify true positives for these alerts, analysts need to scrutinise the context around them. One way of doing this is to enhance the alert by looking up indicators and gathering relevant \emph{cyber threat intelligence}. However, if this is sufficient to conclude, we argue that the alert falls back into the first category -- the alert can be analysed perfectly in isolation. The other way of approaching contextualisation is to look for associated alerts and try to stitch together a picture of what happened\footnote{To use a simple example, alerts for brute-force password attempts towards a user account are a lot more interesting if followed by an alert detailing a new login of the same user from a suspicious location}.

In certain cases, these sequences of corresponding alerts determine patterns that are distinct enough to be hard-coded as multi-alert signatures, and the low-severity alerts can be combined into a single, higher-severity alert. Again, we find that this means that the alert can be moved into the first category. What we are now left with are alerts for which no easy recipes exist, and it is in this difficult domain that we aim to improve analysis.

Even if no absolute recipe is available to analyse an alert as described above, we know something about how human analysts tend to approach them, and we choose to assume that this experience should inform our choice of method. Alerts will normally contain indicators like IP addresses, usernames, file hashes and so forth, and these can often appear in other recent alerts. Choosing properties that seem important and pivoting on them\footnote{\emph{Pivoting} on an indicator means performing a search for other alerts containing the same indicator –- for instance, if you are investigating an alert concerning the host \texttt{web-server01}, pivoting on the hostname means querying for all other alerts having to do with \texttt{web-server01}, normally within some set time frame.} brings context to the original alert. This context can be in the form of low-severity alerts that normally would not be analysed, but can serve as key information when seen together with other alerts.  We are here firmly within the alerts of the third category as defined above.

Pivoting on a single property can be described as asking for the timeline of all alerts involving this property. But it is by pivoting on new alerts, and subsequently performing new pivots on the new properties found in these new alerts, that the full picture sometimes emerges -- in other words, \emph{nested pivoting}.

To formalise the process of nested pivoting and avoid making too many limiting choices, we choose to perform all reasonable pivots on all alerts, creating timelines for all reasonable indicators, and stitching them together into what can best be described as a chronologically ordered graph of alerts. This format is an efficient way of representing alerts as seen from an experienced indicator-pivoting analyst's point of view. With the assumption that their way of approaching alerts is valuable, we can hope that we have created a structure that preserves important information. Said differently, if we believe that the information analysts gather by pivoting is of any value, and we want machines to be able to emulate parts of what they do, bringing this pivoting information into the alert group format is necessary.

% that have the following properties:
% \begin{itemize}
%     \item Represents one step of a larger killchain
%     \item The alerts are connected in a meaningful way
%     \item There is some geometric information
%     \item Alerts are abstracted  
% \end{itemize}

\begin{figure}[!ht]
\centering
\tikzset{every picture/.style={line width=0.75pt}} %set default line width to 0.75pt        

\begin{tikzpicture}[x=0.45pt,y=0.45pt,yscale=-1,xscale=1]
%uncomment if require: \path (0,678); %set diagram left start at 0, and has height of 678

%Shape: Circle [id:dp9862012359233338] 
\draw  [fill={rgb, 255:red, 248; green, 231; blue, 28 }  ,fill opacity=1 ] (206,77.5) .. controls (206,69.22) and (212.72,62.5) .. (221,62.5) .. controls (229.28,62.5) and (236,69.22) .. (236,77.5) .. controls (236,85.78) and (229.28,92.5) .. (221,92.5) .. controls (212.72,92.5) and (206,85.78) .. (206,77.5) -- cycle ;

%Shape: Circle [id:dp6218816883347356] 
\draw  [fill={rgb, 255:red, 184; green, 233; blue, 134 }  ,fill opacity=1 ] (265.43,77.5) .. controls (265.43,69.22) and (272.15,62.5) .. (280.43,62.5) .. controls (288.71,62.5) and (295.43,69.22) .. (295.43,77.5) .. controls (295.43,85.78) and (288.71,92.5) .. (280.43,92.5) .. controls (272.15,92.5) and (265.43,85.78) .. (265.43,77.5) -- cycle ;

%Shape: Circle [id:dp25397606091506075] 
\draw  [fill={rgb, 255:red, 74; green, 144; blue, 226 }  ,fill opacity=1 ] (324.86,77.5) .. controls (324.86,69.22) and (331.58,62.5) .. (339.86,62.5) .. controls (348.14,62.5) and (354.86,69.22) .. (354.86,77.5) .. controls (354.86,85.78) and (348.14,92.5) .. (339.86,92.5) .. controls (331.58,92.5) and (324.86,85.78) .. (324.86,77.5) -- cycle ;

%Shape: Circle [id:dp9999041306583435] 
\draw  [fill={rgb, 255:red, 184; green, 233; blue, 134 }  ,fill opacity=1 ] (384.29,77.5) .. controls (384.29,69.22) and (391.01,62.5) .. (399.29,62.5) .. controls (407.57,62.5) and (414.29,69.22) .. (414.29,77.5) .. controls (414.29,85.78) and (407.57,92.5) .. (399.29,92.5) .. controls (391.01,92.5) and (384.29,85.78) .. (384.29,77.5) -- cycle ;

%Shape: Circle [id:dp20542442137667039] 
\draw  [fill={rgb, 255:red, 248; green, 231; blue, 28 }  ,fill opacity=1 ] (443.72,77.5) .. controls (443.72,69.22) and (450.44,62.5) .. (458.72,62.5) .. controls (467,62.5) and (473.72,69.22) .. (473.72,77.5) .. controls (473.72,85.78) and (467,92.5) .. (458.72,92.5) .. controls (450.44,92.5) and (443.72,85.78) .. (443.72,77.5) -- cycle ;

%Shape: Circle [id:dp7804358331688843] 
\draw  [fill={rgb, 255:red, 245; green, 166; blue, 35 }  ,fill opacity=1 ] (503.15,77.5) .. controls (503.15,69.22) and (509.87,62.5) .. (518.15,62.5) .. controls (526.43,62.5) and (533.15,69.22) .. (533.15,77.5) .. controls (533.15,85.78) and (526.43,92.5) .. (518.15,92.5) .. controls (509.87,92.5) and (503.15,85.78) .. (503.15,77.5) -- cycle ;

%Shape: Circle [id:dp2651572815918959] 
\draw  [fill={rgb, 255:red, 248; green, 231; blue, 28 }  ,fill opacity=1 ] (562.58,77.5) .. controls (562.58,69.22) and (569.3,62.5) .. (577.58,62.5) .. controls (585.86,62.5) and (592.58,69.22) .. (592.58,77.5) .. controls (592.58,85.78) and (585.86,92.5) .. (577.58,92.5) .. controls (569.3,92.5) and (562.58,85.78) .. (562.58,77.5) -- cycle ;

%Shape: Circle [id:dp07545180090615033] 
\draw  [fill={rgb, 255:red, 184; green, 233; blue, 134 }  ,fill opacity=1 ] (622,77.5) .. controls (622,69.22) and (628.72,62.5) .. (637,62.5) .. controls (645.28,62.5) and (652,69.22) .. (652,77.5) .. controls (652,85.78) and (645.28,92.5) .. (637,92.5) .. controls (628.72,92.5) and (622,85.78) .. (622,77.5) -- cycle ;

%Shape: Circle [id:dp5501851317013773] 
\draw  [fill={rgb, 255:red, 248; green, 231; blue, 28 }  ,fill opacity=1 ] (207,513.5) .. controls (207,505.22) and (213.72,498.5) .. (222,498.5) .. controls (230.28,498.5) and (237,505.22) .. (237,513.5) .. controls (237,521.78) and (230.28,528.5) .. (222,528.5) .. controls (213.72,528.5) and (207,521.78) .. (207,513.5) -- cycle ;

%Shape: Circle [id:dp5216770755754018] 
\draw  [fill={rgb, 255:red, 184; green, 233; blue, 134 }  ,fill opacity=1 ] (273.43,470.5) .. controls (273.43,462.22) and (280.15,455.5) .. (288.43,455.5) .. controls (296.71,455.5) and (303.43,462.22) .. (303.43,470.5) .. controls (303.43,478.78) and (296.71,485.5) .. (288.43,485.5) .. controls (280.15,485.5) and (273.43,478.78) .. (273.43,470.5) -- cycle ;

%Shape: Circle [id:dp5954935032448575] 
\draw  [fill={rgb, 255:red, 74; green, 144; blue, 226 }  ,fill opacity=1 ] (308.86,568.5) .. controls (308.86,560.22) and (315.58,553.5) .. (323.86,553.5) .. controls (332.14,553.5) and (338.86,560.22) .. (338.86,568.5) .. controls (338.86,576.78) and (332.14,583.5) .. (323.86,583.5) .. controls (315.58,583.5) and (308.86,576.78) .. (308.86,568.5) -- cycle ;

%Shape: Circle [id:dp47149731111727766] 
\draw  [fill={rgb, 255:red, 184; green, 233; blue, 134 }  ,fill opacity=1 ] (385.29,513.5) .. controls (385.29,505.22) and (392.01,498.5) .. (400.29,498.5) .. controls (408.57,498.5) and (415.29,505.22) .. (415.29,513.5) .. controls (415.29,521.78) and (408.57,528.5) .. (400.29,528.5) .. controls (392.01,528.5) and (385.29,521.78) .. (385.29,513.5) -- cycle ;

%Shape: Circle [id:dp6869513092416866] 
\draw  [fill={rgb, 255:red, 248; green, 231; blue, 28 }  ,fill opacity=1 ] (444.72,513.5) .. controls (444.72,505.22) and (451.44,498.5) .. (459.72,498.5) .. controls (468,498.5) and (474.72,505.22) .. (474.72,513.5) .. controls (474.72,521.78) and (468,528.5) .. (459.72,528.5) .. controls (451.44,528.5) and (444.72,521.78) .. (444.72,513.5) -- cycle ;

%Shape: Circle [id:dp22424082316893235] 
\draw  [fill={rgb, 255:red, 245; green, 166; blue, 35 }  ,fill opacity=1 ] (497.15,568.5) .. controls (497.15,560.22) and (503.87,553.5) .. (512.15,553.5) .. controls (520.43,553.5) and (527.15,560.22) .. (527.15,568.5) .. controls (527.15,576.78) and (520.43,583.5) .. (512.15,583.5) .. controls (503.87,583.5) and (497.15,576.78) .. (497.15,568.5) -- cycle ;

%Shape: Circle [id:dp21951686684306504] 
\draw  [fill={rgb, 255:red, 248; green, 231; blue, 28 }  ,fill opacity=1 ] (563.58,513.5) .. controls (563.58,505.22) and (570.3,498.5) .. (578.58,498.5) .. controls (586.86,498.5) and (593.58,505.22) .. (593.58,513.5) .. controls (593.58,521.78) and (586.86,528.5) .. (578.58,528.5) .. controls (570.3,528.5) and (563.58,521.78) .. (563.58,513.5) -- cycle ;

%Shape: Circle [id:dp38306889650960474] 
\draw  [fill={rgb, 255:red, 184; green, 233; blue, 134 }  ,fill opacity=1 ] (622,566.5) .. controls (622,558.22) and (628.72,551.5) .. (637,551.5) .. controls (645.28,551.5) and (652,558.22) .. (652,566.5) .. controls (652,574.78) and (645.28,581.5) .. (637,581.5) .. controls (628.72,581.5) and (622,574.78) .. (622,566.5) -- cycle ;

%Straight Lines [id:da6635112193615251] 
\draw [line width=1.5]    (233.5,505) .. controls (233.98,502.69) and (235.37,501.78) .. (237.68,502.25) .. controls (239.99,502.72) and (241.38,501.81) .. (241.85,499.5) .. controls (242.33,497.19) and (243.72,496.28) .. (246.03,496.75) .. controls (248.34,497.22) and (249.73,496.31) .. (250.2,494) .. controls (250.68,491.69) and (252.07,490.78) .. (254.38,491.25) .. controls (256.69,491.72) and (258.08,490.81) .. (258.56,488.5) .. controls (259.03,486.19) and (260.42,485.28) .. (262.73,485.75) -- (265.31,484.05) -- (271.99,479.65) ;
\draw [shift={(274.5,478)}, rotate = 146.63] [color={rgb, 255:red, 0; green, 0; blue, 0 }  ][line width=1.5]    (14.21,-4.28) .. controls (9.04,-1.82) and (4.3,-0.39) .. (0,0) .. controls (4.3,0.39) and (9.04,1.82) .. (14.21,4.28)   ;
%Straight Lines [id:da46228031373482237] 
\draw [line width=1.5]    (302.5,477) -- (382.68,505.98) ;
\draw [shift={(385.5,507)}, rotate = 199.87] [color={rgb, 255:red, 0; green, 0; blue, 0 }  ][line width=1.5]    (14.21,-4.28) .. controls (9.04,-1.82) and (4.3,-0.39) .. (0,0) .. controls (4.3,0.39) and (9.04,1.82) .. (14.21,4.28)   ;
%Straight Lines [id:da24318539717775134] 
\draw [line width=1.5]  [dash pattern={on 1.69pt off 2.76pt}]  (335.5,558) -- (385.93,527.55) ;
\draw [shift={(388.5,526)}, rotate = 148.88] [color={rgb, 255:red, 0; green, 0; blue, 0 }  ][line width=1.5]    (14.21,-4.28) .. controls (9.04,-1.82) and (4.3,-0.39) .. (0,0) .. controls (4.3,0.39) and (9.04,1.82) .. (14.21,4.28)   ;
%Straight Lines [id:da5013184361365606] 
\draw [line width=1.5]    (589.5,525) .. controls (591.85,524.76) and (593.14,525.81) .. (593.37,528.16) .. controls (593.61,530.51) and (594.9,531.56) .. (597.25,531.32) .. controls (599.6,531.08) and (600.89,532.13) .. (601.12,534.48) .. controls (601.36,536.83) and (602.65,537.88) .. (605,537.64) .. controls (607.35,537.4) and (608.64,538.45) .. (608.87,540.8) .. controls (609.11,543.15) and (610.4,544.2) .. (612.75,543.96) .. controls (615.1,543.72) and (616.39,544.77) .. (616.62,547.12) -- (618.98,549.05) -- (625.18,554.1) ;
\draw [shift={(627.5,556)}, rotate = 219.21] [color={rgb, 255:red, 0; green, 0; blue, 0 }  ][line width=1.5]    (14.21,-4.28) .. controls (9.04,-1.82) and (4.3,-0.39) .. (0,0) .. controls (4.3,0.39) and (9.04,1.82) .. (14.21,4.28)   ;
%Straight Lines [id:da04866432856022451] 
\draw [line width=1.5]    (413.5,505) -- (444.5,505) ;
\draw [shift={(447.5,505)}, rotate = 180] [color={rgb, 255:red, 0; green, 0; blue, 0 }  ][line width=1.5]    (14.21,-4.28) .. controls (9.04,-1.82) and (4.3,-0.39) .. (0,0) .. controls (4.3,0.39) and (9.04,1.82) .. (14.21,4.28)   ;
%Straight Lines [id:da9786591176349041] 
\draw [line width=1.5]  [dash pattern={on 1.69pt off 2.76pt}]  (474.72,513.5) -- (560.58,513.5) ;
\draw [shift={(563.58,513.5)}, rotate = 180] [color={rgb, 255:red, 0; green, 0; blue, 0 }  ][line width=1.5]    (14.21,-4.28) .. controls (9.04,-1.82) and (4.3,-0.39) .. (0,0) .. controls (4.3,0.39) and (9.04,1.82) .. (14.21,4.28)   ;
%Straight Lines [id:da1132316059333518] 
\draw [line width=1.5]    (471.5,522) -- (619.12,565.65) ;
\draw [shift={(622,566.5)}, rotate = 196.47] [color={rgb, 255:red, 0; green, 0; blue, 0 }  ][line width=1.5]    (14.21,-4.28) .. controls (9.04,-1.82) and (4.3,-0.39) .. (0,0) .. controls (4.3,0.39) and (9.04,1.82) .. (14.21,4.28)   ;
%Straight Lines [id:da806093589864688] 
\draw [line width=1.5]    (338.86,568.5) -- (494.15,568.5) ;
\draw [shift={(497.15,568.5)}, rotate = 180] [color={rgb, 255:red, 0; green, 0; blue, 0 }  ][line width=1.5]    (14.21,-4.28) .. controls (9.04,-1.82) and (4.3,-0.39) .. (0,0) .. controls (4.3,0.39) and (9.04,1.82) .. (14.21,4.28)   ;
%Shape: Circle [id:dp3506298645416033] 
\draw  [fill={rgb, 255:red, 248; green, 231; blue, 28 }  ,fill opacity=1 ] (206,161) .. controls (206,152.72) and (212.72,146) .. (221,146) .. controls (229.28,146) and (236,152.72) .. (236,161) .. controls (236,169.28) and (229.28,176) .. (221,176) .. controls (212.72,176) and (206,169.28) .. (206,161) -- cycle ;

%Straight Lines [id:da7989251843322029] 
\draw [line width=1.5]    (236,161) -- (263.43,161) ;
\draw [shift={(266.43,161)}, rotate = 180] [color={rgb, 255:red, 0; green, 0; blue, 0 }  ][line width=1.5]    (14.21,-4.28) .. controls (9.04,-1.82) and (4.3,-0.39) .. (0,0) .. controls (4.3,0.39) and (9.04,1.82) .. (14.21,4.28)   ;
%Shape: Circle [id:dp5221382727620283] 
\draw  [fill={rgb, 255:red, 184; green, 233; blue, 134 }  ,fill opacity=1 ] (266.43,161) .. controls (266.43,152.72) and (273.15,146) .. (281.43,146) .. controls (289.71,146) and (296.43,152.72) .. (296.43,161) .. controls (296.43,169.28) and (289.71,176) .. (281.43,176) .. controls (273.15,176) and (266.43,169.28) .. (266.43,161) -- cycle ;

%Shape: Circle [id:dp07436858429579363] 
\draw  [fill={rgb, 255:red, 184; green, 233; blue, 134 }  ,fill opacity=1 ] (266.43,218.75) .. controls (266.43,210.47) and (273.15,203.75) .. (281.43,203.75) .. controls (289.71,203.75) and (296.43,210.47) .. (296.43,218.75) .. controls (296.43,227.03) and (289.71,233.75) .. (281.43,233.75) .. controls (273.15,233.75) and (266.43,227.03) .. (266.43,218.75) -- cycle ;

%Shape: Circle [id:dp1521134695287989] 
\draw  [fill={rgb, 255:red, 184; green, 233; blue, 134 }  ,fill opacity=1 ] (385.96,218.75) .. controls (385.96,210.47) and (392.68,203.75) .. (400.96,203.75) .. controls (409.24,203.75) and (415.96,210.47) .. (415.96,218.75) .. controls (415.96,227.03) and (409.24,233.75) .. (400.96,233.75) .. controls (392.68,233.75) and (385.96,227.03) .. (385.96,218.75) -- cycle ;

%Shape: Circle [id:dp8273313062495116] 
\draw  [fill={rgb, 255:red, 248; green, 231; blue, 28 }  ,fill opacity=1 ] (446.56,218.75) .. controls (446.56,210.47) and (453.28,203.75) .. (461.56,203.75) .. controls (469.84,203.75) and (476.56,210.47) .. (476.56,218.75) .. controls (476.56,227.03) and (469.84,233.75) .. (461.56,233.75) .. controls (453.28,233.75) and (446.56,227.03) .. (446.56,218.75) -- cycle ;

%Shape: Circle [id:dp46716909300181897] 
\draw  [fill={rgb, 255:red, 184; green, 233; blue, 134 }  ,fill opacity=1 ] (622,218.75) .. controls (622,210.47) and (628.72,203.75) .. (637,203.75) .. controls (645.28,203.75) and (652,210.47) .. (652,218.75) .. controls (652,227.03) and (645.28,233.75) .. (637,233.75) .. controls (628.72,233.75) and (622,227.03) .. (622,218.75) -- cycle ;

%Straight Lines [id:da46249497192683886] 
\draw [line width=1.5]    (415.96,218.75) -- (443.56,218.75) ;
\draw [shift={(446.56,218.75)}, rotate = 180] [color={rgb, 255:red, 0; green, 0; blue, 0 }  ][line width=1.5]    (14.21,-4.28) .. controls (9.04,-1.82) and (4.3,-0.39) .. (0,0) .. controls (4.3,0.39) and (9.04,1.82) .. (14.21,4.28)   ;
%Straight Lines [id:da0988049958807945] 
\draw [line width=1.5]    (476.56,218.75) -- (619,218.75) ;
\draw [shift={(622,218.75)}, rotate = 180] [color={rgb, 255:red, 0; green, 0; blue, 0 }  ][line width=1.5]    (14.21,-4.28) .. controls (9.04,-1.82) and (4.3,-0.39) .. (0,0) .. controls (4.3,0.39) and (9.04,1.82) .. (14.21,4.28)   ;
%Straight Lines [id:da41406297924803903] 
\draw [line width=1.5]    (296.43,218.75) -- (382.96,218.75) ;
\draw [shift={(385.96,218.75)}, rotate = 180] [color={rgb, 255:red, 0; green, 0; blue, 0 }  ][line width=1.5]    (14.21,-4.28) .. controls (9.04,-1.82) and (4.3,-0.39) .. (0,0) .. controls (4.3,0.39) and (9.04,1.82) .. (14.21,4.28)   ;

%Shape: Circle [id:dp42126121288953045] 
\draw  [fill={rgb, 255:red, 74; green, 144; blue, 226 }  ,fill opacity=1 ] (324.86,276.5) .. controls (324.86,268.22) and (331.58,261.5) .. (339.86,261.5) .. controls (348.14,261.5) and (354.86,268.22) .. (354.86,276.5) .. controls (354.86,284.78) and (348.14,291.5) .. (339.86,291.5) .. controls (331.58,291.5) and (324.86,284.78) .. (324.86,276.5) -- cycle ;

%Shape: Circle [id:dp8681979612021197] 
\draw  [fill={rgb, 255:red, 248; green, 231; blue, 28 }  ,fill opacity=1 ] (559.58,276.5) .. controls (559.58,268.22) and (566.3,261.5) .. (574.58,261.5) .. controls (582.86,261.5) and (589.58,268.22) .. (589.58,276.5) .. controls (589.58,284.78) and (582.86,291.5) .. (574.58,291.5) .. controls (566.3,291.5) and (559.58,284.78) .. (559.58,276.5) -- cycle ;

%Straight Lines [id:da2811577190484994] 
\draw [line width=1.5]    (354.86,276.5) -- (382.96,276.5) ;
\draw [shift={(385.96,276.5)}, rotate = 180] [color={rgb, 255:red, 0; green, 0; blue, 0 }  ][line width=1.5]    (14.21,-4.28) .. controls (9.04,-1.82) and (4.3,-0.39) .. (0,0) .. controls (4.3,0.39) and (9.04,1.82) .. (14.21,4.28)   ;
%Straight Lines [id:da9261063312366146] 
\draw [line width=1.5]    (476.56,276.5) -- (554.58,276.5) ;
\draw [shift={(557.58,276.5)}, rotate = 180] [color={rgb, 255:red, 0; green, 0; blue, 0 }  ][line width=1.5]    (14.21,-4.28) .. controls (9.04,-1.82) and (4.3,-0.39) .. (0,0) .. controls (4.3,0.39) and (9.04,1.82) .. (14.21,4.28)   ;
%Shape: Circle [id:dp41650971070542264] 
\draw  [fill={rgb, 255:red, 184; green, 233; blue, 134 }  ,fill opacity=1 ] (385.96,276.5) .. controls (385.96,268.22) and (392.68,261.5) .. (400.96,261.5) .. controls (409.24,261.5) and (415.96,268.22) .. (415.96,276.5) .. controls (415.96,284.78) and (409.24,291.5) .. (400.96,291.5) .. controls (392.68,291.5) and (385.96,284.78) .. (385.96,276.5) -- cycle ;

%Shape: Circle [id:dp14904062410325736] 
\draw  [fill={rgb, 255:red, 248; green, 231; blue, 28 }  ,fill opacity=1 ] (446.56,276.5) .. controls (446.56,268.22) and (453.28,261.5) .. (461.56,261.5) .. controls (469.84,261.5) and (476.56,268.22) .. (476.56,276.5) .. controls (476.56,284.78) and (469.84,291.5) .. (461.56,291.5) .. controls (453.28,291.5) and (446.56,284.78) .. (446.56,276.5) -- cycle ;

%Straight Lines [id:da25115726646604175] 
\draw [line width=1.5]    (415.96,276.5) -- (443.56,276.5) ;
\draw [shift={(446.56,276.5)}, rotate = 180] [color={rgb, 255:red, 0; green, 0; blue, 0 }  ][line width=1.5]    (14.21,-4.28) .. controls (9.04,-1.82) and (4.3,-0.39) .. (0,0) .. controls (4.3,0.39) and (9.04,1.82) .. (14.21,4.28)   ;

%Shape: Circle [id:dp8342679415821111] 
\draw  [fill={rgb, 255:red, 245; green, 166; blue, 35 }  ,fill opacity=1 ] (500.15,334.25) .. controls (500.15,325.97) and (506.87,319.25) .. (515.15,319.25) .. controls (523.43,319.25) and (530.15,325.97) .. (530.15,334.25) .. controls (530.15,342.53) and (523.43,349.25) .. (515.15,349.25) .. controls (506.87,349.25) and (500.15,342.53) .. (500.15,334.25) -- cycle ;

%Straight Lines [id:da19489561673594835] 
\draw [line width=1.5]    (354.86,334.25) -- (497.15,333.76) ;
\draw [shift={(500.15,333.75)}, rotate = 179.8] [color={rgb, 255:red, 0; green, 0; blue, 0 }  ][line width=1.5]    (14.21,-4.28) .. controls (9.04,-1.82) and (4.3,-0.39) .. (0,0) .. controls (4.3,0.39) and (9.04,1.82) .. (14.21,4.28)   ;
%Shape: Circle [id:dp4732609420335627] 
\draw  [fill={rgb, 255:red, 74; green, 144; blue, 226 }  ,fill opacity=1 ] (324.86,334.25) .. controls (324.86,325.97) and (331.58,319.25) .. (339.86,319.25) .. controls (348.14,319.25) and (354.86,325.97) .. (354.86,334.25) .. controls (354.86,342.53) and (348.14,349.25) .. (339.86,349.25) .. controls (331.58,349.25) and (324.86,342.53) .. (324.86,334.25) -- cycle ;

%Straight Lines [id:da12066924444232097] 
\draw [line width=1.5]    (595.58,392) -- (621,392) ;
\draw [shift={(624,392)}, rotate = 180] [color={rgb, 255:red, 0; green, 0; blue, 0 }  ][line width=1.5]    (14.21,-4.28) .. controls (9.04,-1.82) and (4.3,-0.39) .. (0,0) .. controls (4.3,0.39) and (9.04,1.82) .. (14.21,4.28)   ;
%Shape: Circle [id:dp30385569509720567] 
\draw  [fill={rgb, 255:red, 248; green, 231; blue, 28 }  ,fill opacity=1 ] (565.58,392) .. controls (565.58,383.72) and (572.3,377) .. (580.58,377) .. controls (588.86,377) and (595.58,383.72) .. (595.58,392) .. controls (595.58,400.28) and (588.86,407) .. (580.58,407) .. controls (572.3,407) and (565.58,400.28) .. (565.58,392) -- cycle ;

%Shape: Circle [id:dp8780035105954853] 
\draw  [fill={rgb, 255:red, 184; green, 233; blue, 134 }  ,fill opacity=1 ] (624,392) .. controls (624,383.72) and (630.72,377) .. (639,377) .. controls (647.28,377) and (654,383.72) .. (654,392) .. controls (654,400.28) and (647.28,407) .. (639,407) .. controls (630.72,407) and (624,400.28) .. (624,392) -- cycle ;

%Straight Lines [id:da3608021187418403] 
\draw [line width=1.5]  [dash pattern={on 1.69pt off 2.76pt}]  (415.5,519) -- (442.5,519) ;
\draw [shift={(445.5,519)}, rotate = 180] [color={rgb, 255:red, 0; green, 0; blue, 0 }  ][line width=1.5]    (14.21,-4.28) .. controls (9.04,-1.82) and (4.3,-0.39) .. (0,0) .. controls (4.3,0.39) and (9.04,1.82) .. (14.21,4.28)   ;
%Shape: Rectangle [id:dp28804497119837547] 
\draw   (190,38) -- (674,38) -- (674,111) -- (190,111) -- cycle ;
%Shape: Rectangle [id:dp9328956602310801] 
\draw   (191.5,135) -- (672.5,135) -- (672.5,419) -- (191.5,419) -- cycle ;
%Shape: Rectangle [id:dp6392831563787709] 
\draw   (190,446) -- (674,446) -- (674,603) -- (190,603) -- cycle ;

% Text Nodes - Modified for centering
\node at (221, 77.5) {A};
\node at (280.43, 77.5) {B};
\node at (339.86, 77.5) {C};
\node at (399.29, 77.5) {D};
\node at (458.72, 77.5) {E};
\node at (518.15, 77.5) {F};
\node at (577.58, 77.5) {G};
\node at (637, 77.5) {H};

\node at (222, 513.5) {A};
\node at (288.43, 470.5) {B};
\node at (323.86, 568.5) {C};
\node at (400.29, 513.5) {D};
\node at (459.72, 513.5) {E};
\node at (512.15, 568.5) {F};
\node at (578.58, 513.5) {G};
\node at (637, 566.5) {H};

\node at (221, 161) {A};
\node at (281.43, 161) {B}; % Circle dp5221382727620283
\node at (281.43, 218.75) {B}; % Circle dp07436858429579363
\node at (339.86, 276.5) {C}; % Circle dp42126121288953045
\node at (339.86, 334.25) {C}; % Circle dp4732609420335627
\node at (400.96, 218.75) {D}; % Circle dp1521134695287989
\node at (400.96, 276.5) {D}; % Circle dp41650971070542264
\node at (461.56, 218.75) {E}; % Circle dp8273313062495116
\node at (461.56, 276.5) {E}; % Circle dp14904062410325736
\node at (515.15, 334.25) {F}; % Circle dp8342679415821111
\node at (574.58, 276.5) {G}; % Circle dp8681979612021197 (Note: original text was (568.58,267.5) for G, circle center (574.58,276.5))
\node at (580.58, 392) {G}; % Circle dp30385569509720567
\node at (637, 218.75) {H}; % Circle dp46716909300181897
\node at (639, 392) {H}; % Circle dp8780035105954853

\end{tikzpicture}
\caption{Top: Alerts ordered chronologically, coloured by type. Middle: Timeline-defining properties have been extracted from the alerts, and the ones that share e.g. a hostname have been placed on the same timeline, revealing some relationships. Bottom: Every timeline-defining property has been taken into account, and we therefore get an alert graph. Now, alert $C$ and alert $B$ can end up in the same group, which is sometimes helpful if they both contributed to what triggered alert $D$. Notice that two properties connect alert $D$ and $E$, which we have drawn as two edges, but in the structure, we use a single edge with features. Deleting edges that represent long time differences leads to alert groups.}
\label{fig:alerts_to_alert_graphs}
\end{figure}
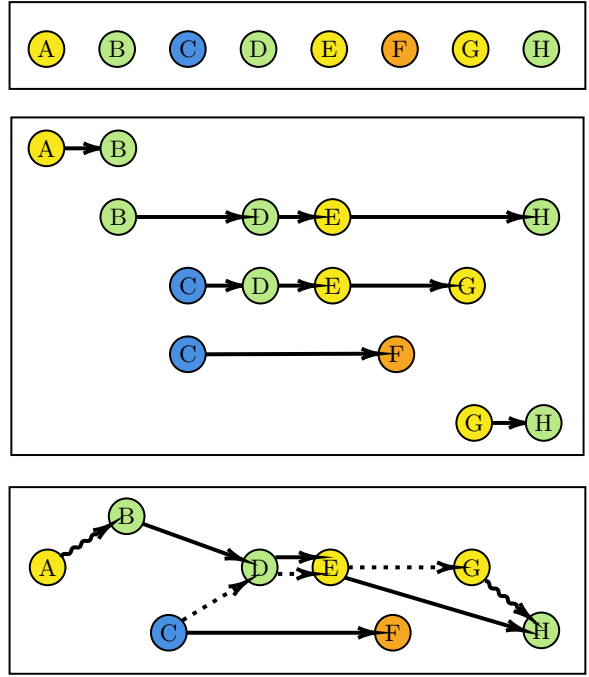

\subsection{Building the alert graph}

To define an alert graph, we first cover the static case where we are given a sequence of alerts $A = (A_1, A_2,..., A_n)$, which we, without loss of generality, will assume to be chronologically ordered. They will be the nodes of the alert graph $G$. We also need to define a set of objects $\tau$ we want to use for building edges in the graph; we call them timeline-defining properties. For now, let these objects be of a single type, like IP addresses, so that $\tau$ is the set of all valid IP addresses.

For every alert in $A$, we then extract its associated IP addresses, meaning that we associate a subset $T_i$ of $\tau$ to every $A_i$. This subset can be empty, and in this case the alert will remain an isolated node in the alert graph. The algorithm for extracting the sets $T_i$ needs to be decided on individually for every alert type involved. We denote the set of IP addresses that appear in our alerts by $\tau' = \bigcup T_i$ (note that this subset of $\tau$ could change whenever a new alert is added, if working on a stream of alerts).

Next, for every individual IP address $p$ in $\tau'$, we create a timeline along it, by linking every consecutive alert $A_i$ where $p\in T_i$ with an edge. To preserve information about what linked the two alerts, we store the value $p$ as an edge feature for the new edge. To preserve chronological details, we make the edge directed and weighted by the time difference between the alerts it sits between. We do not directly link all alerts that share the IP address $p$. They all end up on a linear graph, ordered by a timestamp\footnote{In some detection systems, alerts with the same timestamp can be a common occurrence. In this case, we make an arbitrary choice of how to order the alerts with the same timestamp, and the resulting alert groups will not be affected by the choice.}. Doing this for every IP address in $\tau'$ results in an alert graph that is essentially a union of intertwined alert timelines, one for each relevant IP address. If two consecutive alerts have more than one IP address in common, we should link them by one edge for each instance, separable by their different edge feature values. To avoid working with multi-graphs, we prefer to link by a single edge that has all the relevant IP address values as features.

In the more general case, we define multiple alert properties that are used for linking, as long as we can define a way of extracting them from the alerts. Later in this paper, hostnames and usernames are used in addition to IP addresses, but any property that analysts find valuable for pivoting between alerts should be considered. When adding features to the edges in this general case, we have to add both the property type(s) and the property value(s). Given $k$ different property types chosen as timeline-defining properties, we write $\tau = \tau_1\cup...\cup\tau_k$ for the complete set.\footnote{
% It is tempting to throw all timeline-defining properties into the same set, but we want to be sure that we are able to separate between special cases like both a username and a hostname being equal to ``\texttt{abc}'' (in which case there should not be an edge between two alerts if they refer to two different objects called ``\texttt{abc}'' ). 
To ``featurise'' the edges correctly, we will think of the timeline-defining properties as formally looking like tuples e.g. $(\texttt{abc}, \text{hostname})$ and $(\texttt{abc}, \text{username})$.}

The alert graph can be considered completed, with all alert information gathered in the corresponding nodes. Alerts tend to be represented using JSON or similar formats, but for most machine learning applications we need a featurisation of these alert properties into Euclidean space. After a vectorisation algorithm $v:A\to\mathbb{R}^n$ is decided on, we replace the full alert information $A_i$ in a node with the node vector $v(A_i)$.
In the end, we arrive at an alert graph $G$ which is a directed acyclic graph (DAG) with multi-attributed edges and nodes.

In practice, we want the alert graph to be a lightweight format suited for a fast stream of incoming alerts, so its actual computation is a little different from that described above. The graph is initialised as an empty graph. In chronological order, incoming alerts are sent through the two functions that extract their timeline-defining properties and their node vector. Next, the timeline-defining properties are checked against a list of current timeline-defining properties in the graph, which is actually a list of keys for a dictionary mapping the properties to the most recent graph node sharing that property.\footnote{In an actual environment, there may be specific property values, like the IP address of a proxy server, that show up in so many alerts that they tie together systems for no good reason. In these cases, blacklisting specific values so that they do not create edges in the graph is a solution.} For every match, a correspondingly featurised edge is added between the new alert node and the matching node, and the new node takes over the place in the dictionary as head of the timeline for that property. We detail the algorithm with pseudocode in \autoref{algo:graph_creation}.

\begin{algorithm}[h]
    \caption{Alert Graph Creation}
    \label{algo:graph_creation}
    \begin{algorithmic}%[1] % The [1] argument instructs to number the lines
    
    \Procedure{GenerateAlertGraph}{$A, \tau, \nu, \Delta$}
        \For{$A_i$ in $A$}
            \State $timelineProperties \gets e(A_i)$
            \State $newNodeVector \gets$ $v(A_i)$
            \State addNode($alertGraph$, $newNodeVector$, $newNodeId = i$)
            \For{$(property, propertyType)$ in $timelineProperties$}
                \If {$(property, propertyType)\in\tau'$}
                    \State $pred \gets headOfTimelinesDict[(property, propertyType)]$
                    \State addEdge($alertGraph, pred, newNodeId, edgeFeatures$)
                    \State $headOfTimelinesDict[(property, propertyType)] \gets newNodeId$
                \Else
                    \State $headOfTimelinesDict[(property, propertyType)] \gets newNodeId$
                \EndIf
            \EndFor
        \EndFor
    \EndProcedure
    
    \end{algorithmic}
\end{algorithm}

\subsection{Creating graph-based alert groups }
\label{sub:delta}

The alert graph structure constructed in the previous section ties the alerts from a stream together, which we claim is a natural structure for further analysis. One way of using it when analysing an alert stream is to visualise an alert with its immediate graph neighbourhood, essentially the same as pre-performing all relevant pivots for the alert. However, we also believe that the alert graph structure provides this information robustly, which means that we can use the format for machine-assisted analysis methods.

We use the term \emph{global alert graph} for the graph containing all alerts.
The global alert graph tends to grow very large in large environments. In many real-world cases, it will consist mainly of noise or false positives, simply because most logged events are non-malicious. We are interested in improving the understanding of single alerts in the graph represented by nodes through their surrounding context. If the timeline-defining properties are chosen correctly, this context corresponds to the graph neighbourhood around each alert. One natural machine-enhanced approach would be a node-classification method, which in a featurised graph utilises message-passing techniques to update the information in each node using the surrounding nodes~\cite{PracticalTutorialOnGNNs}.

Another approach is to see clusters of alerts in the graph as coherent groups having to do with the same incident, and consider the longer edges to be links between different events on the same system\footnote{Recall that we included the time-delta between two linked alerts as a feature of the edge, so talking about edge length is meaningful.}. The underlying assumption is that when something suspicious happens in our systems, we have detectors that pick up traces of the activity that result in alerts spanning a relatively short time window. In this case, it can be more natural to consider the global alert graph as a structure from which smaller alert groups should be isolated (or cut off), by removing the longer edges representing links between different events on the same system. Looking at how analysts use properties to pivot between alerts, we are essentially asking how far back in time they would normally look along a hostname or an IP address.

The easiest way to proceed is to choose a delta cut-off time $\delta$ and remove, or never construct, any edge between alerts that have a chronological distance higher than $\delta$. Since we are working with featurised edges, in which each edge contains at least one timeline defining property type, we can also pick a collection of deltas $\Delta = (\delta_1, ..., \delta_k)$ corresponding to the property types $\tau = \tau_1\cup...\cup\tau_k$, and then only keep an edge if it is of type $\tau_j$ and is shorter than $\delta_j$. Given the choice $\Delta$, we immediately go from a global, ever-growing alert graph $G$ to \emph{graph-based alert groups}\footnote{Referred to as `alert groups' throughout this paper for brevity.}. Note that these groups resulting from removing certain edges retain their own graph structure as subgraphs of $G$. The hypothesis is that graph-based alert groups created with suitable timeline-defining properties $\tau$ and time deltas $\delta_j$ can capture single attack steps. In the following section, we will explore a graph learning approach to identifying the related attack step for a given graph-based alert group.

% An interesting -- although not very surprising -- outcome of our experiments is that this set $\Delta$ cannot be chosen universally: For picking up different types of malicious activity, different choices of $\Delta$ give the best alert groups. Methods for choosing, and continuously tuning, the $\Delta$ parameters in an alert system is part of our ongoing research.

% different attacks = different deltas 
% Different steps of cyber attacks are different in nature, and manifest differently in the alerts. For example, a scanning technique might trigger alerts with a time interval of a few seconds as it rapidly probes a network. In contrast, a data exfiltration technique could generate alerts minutes or hours apart, reflecting the slower, stealthier process of extracting sensitive information. 
% We expect different attacks to be best represented in alert graphs with different $\delta$-time cutoffs. Building on this, we also expect that for a given attack step it is optimal to have different $\delta$ for each of the edge types in $\tau$. 

\FloatBarrier

\section{Matching graph-based alert groups}
\label{sec:GMN}
% introduction - like an abstract

%PETERS VARIANT
One of our primary motivations for grouping alerts is to be able to match our generated graph-based alert groups to similar historical graph-based alert groups.
This means that the alert group can be contextualised with data present for the historic event, such as relevant CTI, how the incident was handled and any conclusions from the previous incident handling. This provides additional armoury for the analyst's toolbox. It is also a way of capturing patterns that keep repeating in the environment, or odd similarities that appear across systems over time. The alert groups we generate can be stored in a knowledge base -- all of them or just the ones we deem important. Thus, deciding how to match against new groups is the next challenge.

Having turned our alert stream into alert groups with an underlying graph structure, the task of alert group matching amounts to comparing directed acyclic graphs with featurised nodes and edges. There is a rich literature of methods for doing this. Still, we have distinctive requirements for our graph matching: (1)~the content of the alerts is likely to be the main information contained in an alert group, meaning that the node features should have a high weight; (2)~noise arising from false positives needs to be taken into account, as well as some randomness in the chronology of the alerts, and therefore strict graph-comparison schemes that look for exact substructures are avoided; (3)~alert group size is also expected to vary, even between good matches; (4)~the edge information should contribute as it is also structured data. One promising \emph{Graph Neural Network}(GNN) approach that satisfies these requirements is \emph{Graph Matching Networks} (GMNs) \cite{li2019graph}, which we explore next.

\subsection{The Graph Matching Network approach}

A GMN is a supervised graph neural network that learns the similarity of structured graph objects~\cite{li2019graph}. 
Given a pair of graphs,  a function to compute a similarity score between them is learned. The model uses message passing within and between graphs to create vector space embeddings, employing a cross-graph attention mechanism to compute contextual matching vectors, enabling distance measurement in a shared vector space.

% GMN details in this experiment
We utilise a GMN to learn a similarity measure between alert groups with the goal of correlating incoming (unknown) alert groups to previously seen (known) alert groups stored in a knowledge base.
% An overview of the training phase is shown in \autoref{fig:gmn-training}.
As this is a supervised learning technique, we create pairs of alert groups $<G_1, G_2>$ from a training set, and label them with $L$ as positive if they relate to the same attack step and negative if they relate to different attack steps or to false positives. This forms the tuple $<G_1, G_2, L>$.
In addition to the trained model, a knowledge base of \textit{known attack groups} is created, containing alert groups recorded from previous attacks. The alert groups in the knowledge base may include information to contextualise the new and unknown alert group. Such information could, for instance, be CTI or details about the phase of the attack.
% % ✂️
% \begin{figure}[h]
%     \centering
%     \includegraphics[width=0.7\linewidth]{figures/train.png}
%     \caption{GMN Training}
%     \label{fig:gmn-training}
% \end{figure}

% An overview of the inference phase is shown in \autoref{fig:gmn-inference}.
During the inference phase, the goal is to identify whether an incoming alert group is similar to anything seen before. The alert group is compared with all alert groups in the knowledge base, one by one. The trained GMN measures distances, and if the distance is sufficiently short, the similarity score and related information from the known alert group are returned and can be used for further analysis. If the incoming group is not close to any known groups in the knowledge base, it is an indication that the group is irrelevant, unseen, or part of a novel type of behaviour.
The strength of the cross-graph attention edges in a GMN indicates which nodes (between the two alert groups) that are deemed the most similar~\cite{li2019graph}. This provides an element of explainability that is considered essential when using AI in SOCs \cite{enisaAI23,miloslavskaya_analysis_2018}. Next, we demonstrate both the generation of graph-based alert groups and the use of GMNs to match such groups.

% % ✂️
% \begin{figure}[h]
%     \centering
%     \includegraphics[width=0.7\linewidth]{figures/inference.png}
%     \caption{GMN inference}
%     \label{fig:gmn-inference}
% \end{figure}

\section{Evaluation}
\label{sec:evaluation}

Here, we provide experimental evidence for our graph-based approach to alert groups. We demonstrate and evaluate both the graph-based alert group creation and the subsequent graph correlation, thus addressing all three research questions from section \ref{sec:intro}.
% The evaluation focuses on the following key aspects:
The section consists of the following three parts:
\begin{enumerate}
    \item \emph{Dataset and experimental setup.} We first describe the dataset and the parameters for creating graph-based alert groups. 
    \item \emph{Graph-based alert group creation.} We assess the quality of the graph creation technique by measuring the cluster purity and the silhouette score.
    \item \emph{Graph-based alert group correlation.} Finally, we evaluate the performance of the graph matching approach, by measuring the similarity to historical alert groups.
    %    in correctly correlating current alert groups with historical groups. We focus primarily on accuracy.  
\end{enumerate}

% The goal of experimentation (and how we are testing)
    % for graphs
    % for matching
% The section contains

\subsection{Dataset }
% about the dataset
We use the \emph{AIT Log Data Set V2.0}~\cite{landauer2022maintainable} as our base system logs, as it contains varied log sources enabling different types of intrusion detection systems.  

This dataset contains logs collected in a testbed representing a small enterprise with multiple attack steps. The dataset consists of eight repeated simulations with slight variations in both normal and malicious traffic.
Crucially for our work, this dataset consists of a set of alerts generated from different intrusion detection systems, which in this case is: Suricata, Wazuh, AMiner~\cite{landauer2024introducing} and Telosian~\cite{Antonio2025telosian}. 
This means the alerts are raised from both signature-based and anomaly-based IDSs. 
The resulting data set consists of $2,655,821$ labelled alerts, with the distribution 
of alert labels shown in  \autoref{tab:attack_methods}. The label "-" corresponds to false positive alerts. Most alerts are false positives, mirroring the challenges SOCs face today~\cite{alahmadi202299}. We also observe that the classes are very unbalanced. This is a natural consequence of different attack methods manifesting differently in the alerts. A directory scan (\texttt{dirb}) might request hundreds of thousands of endpoints, while remote code execution (\texttt{webshell\_cmd}) might only be a few actions. 

\begin{table}[h]
    \centering
    \begin{tabular}{|l|r|r|r|r|r|}
        \hline
        \textbf{Attack Method}     & \textbf{Wazuh} & \textbf{Suricata} & \textbf{AMiner} & \textbf{Telosian} & \textbf{Total} \\
        \hline
        dirb                      & 1,670,616     & -             & 18,329  &   6944    & 1,695,889     \\
        \hline
        -                         & 595,604       & 233,523       & 9,444   &    66609    & 905,180       \\
        \hline
        dnsteal                   & -             & 72,890        & 44      &    493     & 73,427        \\
        \hline
        wpscan                    & 27,288        & -             & 27,491  &     146    & 54,922        \\
        \hline
        service\_scan             & 42            & 216           & -       &      2      & 260           \\
        \hline
        escalated\_sudo\_command  & 56            & -             & 127     &      -      & 183           \\
        \hline
        attacker\_change\_user    & 22            & -             & 78      &      -      & 100           \\
        \hline
        webshell\_cmd             & -             & -             & 25      &      -      & 25            \\
        \hline
        crack\_passwords          & -             & -             & 11      &      -      & 11            \\
        \hline
        dns\_scan                 & -             & -             & 9       &      -      & 9             \\
        \hline
        online\_cracking          & -             & 6             & -       &     -       & 6             \\
        \hline
    \end{tabular}
    \caption{Distribution of Alert Labels.}
    \label{tab:attack_methods}
\end{table}

\vspace{-10pt}
From the alerts, we create graph-based alert groups using the timeline-defining properties \texttt{username} and \texttt{IP}. 
As discussed in \autoref{sec:alert_structure}, different $\Delta$ are suitable for representing different attack steps.
The $\Delta$  for the different attack steps are shown in \autoref{tab:time-deltas}.

\begin{table}[h]
    \centering
    \begin{tabular}{|l|r|r|}
        \hline
        \textbf{Attack Method}     & \textbf{Username (seconds)} & \textbf{IP (seconds)}  \\
        \hline
        wpscan                    & 0             & 3               \\
        \hline
        service\_scan             & 0             & 30              \\
        \hline
        escalated\_sudo\_command  & 30            & 3               \\
        \hline
        attacker\_change\_user    & 30            & 3               \\
        \hline
        webshell\_cmd             & 0             & 100             \\
        \hline
    \end{tabular}
    \caption{Time $\Delta$  for different attack steps}
    \label{tab:time-deltas}
\end{table}

% % ✂️
% \begin{figure}[h]
%     \centering
%     \begin{subfigure}{0.45\linewidth}
%         \centering
%         \includegraphics[width=\linewidth]{figures/example_graph.png}
%     \end{subfigure}
%     \hfill
%     \begin{subfigure}{0.45\linewidth}
%         \centering
%         \includegraphics[width=\linewidth]{figures/example_graph_2.png}
%     \end{subfigure}
%     \caption{Graph based alert group visualized}
%     \label{fig:graph-examples}
% \end{figure}

\subsection{Graph-based alert group evaluation}

This first experiment aims to evaluate the creation of graph-based alert groups, addressing the first two research questions (RQ1 and RQ2).
% Show AInception WP4 results but in a concise way
The metrics we deemed relevant are \emph{cluster purity}~\cite{reddy2013data} and \emph{silhouette score}~\cite{Shutaywi2021Silhouette}. 
% purity
The purity of an alert group is the proportion of alerts belonging to the true attack step.
% silhouette score
The silhouette score measures how well an alert group fits within its assigned cluster compared to all other clusters, and can be seen as a measure of the quality of the clustering.
It is calculated by comparing the average distance to all other graphs in the same cluster (intra-cluster distance) with the average distance to graphs in the nearest different cluster (nearest-cluster distance). 
It can thus indicate how recognisable alert groups with the same attack type are by measuring the cohesion within the same cluster and how distinguished they are from other attack groups by measuring their separation. 
% what the scores mean
Determining the silhouette score requires featurisation on the graph level, which is achieved by aggregating the features of the nodes in each graph. The distance is calculated using the Euclidean distance.

% how we tested
The graph-based alert aggregation method is run on all alerts in the data set for each scenario separately using the $\Delta$ outlined in \autoref{tab:time-deltas}. 
This method is compared to a purely time-based alert aggregation method\footnote{A published approach to alert aggregation\cite{landauer2022dealing}}, where alerts are combined if the time between them is shorter than the defined $\Delta$.
It is important to create good groups that capture actual attack steps, while the quality of groups that capture false positive alerts is pointless as they should be filtered out later. For this reason, we only analyse groups capturing actual attack steps. The results are shown in \autoref{tab:grouping_metrics}.

\begin{table}[h!]
\centering
\begin{tabular}{|l|c|c|c|c|}
\hline
\textbf{Metric} & \multicolumn{2}{c|}{\textbf{Standard}} & \multicolumn{2}{c|}{\textbf{Excluding dnsteal}} \\
\cline{2-5}
 & \textbf{Graph-based} & \textbf{Time-based} & \textbf{Graph-based} & \textbf{Time-based} \\
\hline
Silhouette score & 0.97385 & 0.90048 & 0.44622 & 0.06470 \\
\hline
Purity           & 0.81632 & 0.30353 & 0.99850 & 0.95653 \\
\hline
\end{tabular}
\caption{Graph-based VS Time-based alert aggregation}
\label{tab:grouping_metrics}
\end{table}
\vspace{-10pt}

Looking at the first two columns, the silhouette score is very high for both methods, indicating clearly separated clusters, meaning that each group with the same attack type is clustered densely together.
The graph-based method slightly outperforms a pure time-based aggregation. The purity of the graph-based method is high at 81.63\%, indicating little noise, outperforming the time-based method.

Note that \texttt{dnsteal} has many times more groups than any other attack. This will have an extra large influence on the silhouette score. Therefore, it is also interesting to check the recognisability of groups excluding \texttt{dnsteal}; This is shown in the last two columns of \autoref{tab:grouping_metrics}. 
Here, the silhouette score indicates that the graph-based method's clustering abilities are solid. The time-based method created rather poorer groups that led to overlapping clusters, with overlap between graphs from different attack types. This makes it difficult to separate graphs using machine learning models. The purity of both methods is high, with the graph-based method consisting of nearly all the same attack types in each group.

Our initial experiment has shown that graph-based alert aggregation performs well in combining alerts from the same attack step while minimising noise, compared to a more straightforward purely time-based method. The alert groups are purer, recognisable, well-separated, and characteristic for each attack type. In contrast, the time-based method often contains more noise, combines more different attacks in a single group, and has more overlap between alert groups with different attack types. The alert groups show promise for use in machine learning models, since they mainly group alerts from the same attack steps while minimising the noise level. %They are theoretically more recognisable than a simple time-based aggregation.

\subsection{Graph-based alert group matching evaluation }

Our second experiment aims to evaluate how accurately the GMN model can match an alert group (representing an attack step) to previously seen alert groups of the same attack step, while minimising matches with unrelated graphs. This experiment addresses the third research question (RQ3).
% This section seeks to answer the following research question:
% \begin{description}
%     \item[\textbf{RQ3:}] To which degree can Graph Matching Networks correlate current graph based alert groups to related historical incidents?
% \end{description}

% encoding of nodes
\sloppy
We create an encoding of each node as a vector of size $133$. We wish to include as much information from the alerts as possible. 
All features are encoded except for \texttt{IP} and \texttt{User} as we wish to capture behavioural patterns, not identify specific users. We one-hot encode and multi-label encode the categorical features like  \texttt{rule\_groups}. 
Edges are encoded as a vector of size two, reflecting if the edge comes from an \texttt{IP} match, a \texttt{user} match or both.
The distance function used is the dot product. The Adam optimiser~\cite{diederik2014adam} is used for optimising with a learning rate of $0.0001$ to avoid overshooting.
\sloppy
% % NN size
% % NN activation functions
% The graph matching network consists of the following parts:
% \begin{itemize}
%     \item Encoder:
%     \begin{itemize}
%         \item MLP (input 133, output 32)
%         \item MLP (input 2, output 16)
%     \end{itemize}
%     \item Graph Aggregator:
%     \begin{itemize}
%         \item MLP (input 32, output 256)
%         \item MLP (input 128, output 128)
%     \end{itemize}
%     \item Graph Propagation layers (5 total)
%     \begin{itemize}
%         \item sequential message nets (input 80, hidden 64, output 64)
%         \begin{itemize}
%             \item ReLU activation funciton
%         \end{itemize}
%         \item sequential reverse message net (input 80, hidden 64, output 64)
%         \begin{itemize}
%             \item ReLU activation funciton
%         \end{itemize}
%         \item GRU (input 96, output 32)
%     \end{itemize}
% \end{itemize}

% Similarity function
% Treshold value

% Train evaluate details
% Dette er tidkrevende for en SOC også, så det erfordel at det funker for få grafer
% There is little labeled data
One of the challenges for supervised machine learning in a real-world SOC environment is the scarcity of labelled data. 
% most data is benign
This is made worse by the fact that most of the available data is benign or noise.
However, it is reasonable to assume that an SOC stores details of handled incidents and associated alerts. Thus, this can be the source of a small number of high-quality manually labelled alerts. 
This is reflected in our dataset, as each attack step only has a few representative alert groups per simulation.  
On the other hand, unrelated alert groups are more widely available.
% About the test set and metodology
In order to perform a meaningful evaluation of the technique, we do not try to classify attack labels with too few alerts, to avoid evaluating a graph-based approach on graphs with only one node.
We also want to avoid attacks that manifest in massive graphs like \texttt{dirb} and \texttt{dnsteal}. These are scanning attacks that typically result in graphs with more than $10,000$ nodes where statistical methods can be sufficient for analysing them.   
We therefore focus on creating models for the following labels: \texttt{attacker\_change\_user}, \texttt{escalated\_sudo\_command} and \texttt{wpscan}.

% training set
The training set for each label consists of $80\%$ of the relevant alert groups and a random selection of unrelated alert groups, both benign and other attack steps. 
The training stage consists of creating all possible pairs of the relevant alert groups for a given label \textit{<attack, attack>} (positive pairs), along with the same number of pairs of different labels \textit{<attack, benign>} or \textit{<attack, different attack>} (negative pairs). 
We oversample the positive pairs to avoid the model favouring the negative pairs.
% test set
The test set consists of the remaining $20\%$ relevant alert groups representing the attack step, and all alert groups from one of the eight simulations representing unrelated groups\footnote{This is the simulation called `harrison' in \cite{landauer2022maintainable}.}. The knowledge base consists of the relevant alert groups from the training set. This is similar to what would be done in a SOC, where the known previous incidents would be used first to train the GMN and secondly added to the knowledge base.
% What we evaluate
We predict the distance between all alert groups in the knowledge base and all alert groups in the test set. This measures both how well the GMN correlates alert graphs that represent the same attack step, as well as how well the GMN does not correlate alert groups representing the attack step with unrelated alert groups.

A separate model is trained for each of the three attack steps: \texttt{attacker\_change\_user}, \texttt{escalated\_sudo\_command} and \texttt{wpscan}. The model is trained for $450$ epochs, where each batch consists of two labelled pairs: one positive and one negative. 

\begin{figure}
    \centering
    \begin{subfigure}{0.50\linewidth}
        \includegraphics[width=\linewidth]{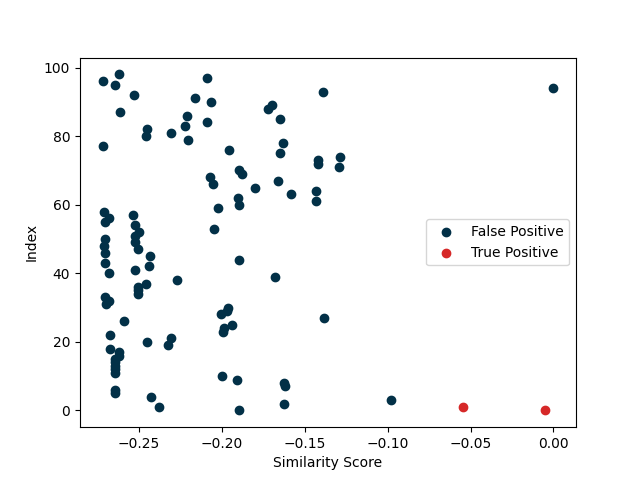}
        \caption{escalated\_sudo\_command}
    \end{subfigure}
    \begin{subfigure}{0.45\linewidth}
        \includegraphics[width=\linewidth]{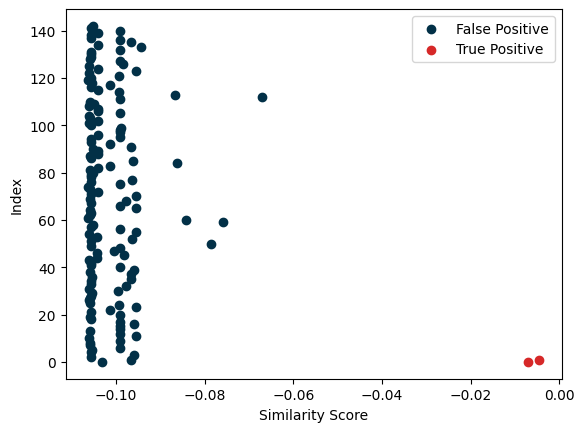}
        \caption{attacker\_change\_user}
    \end{subfigure}
    \caption{Predicted distances between all graphs in 'harrison' dataset and the alert groups in knowledge base for the two attacks}
    \label{fig:pred_closest}
\end{figure}

The columns for \texttt{attacker\_change\_user} and \texttt{escalated\_sudo\_command} in \autoref{tab:GMN_results} show that both labels for the positive pairs have a mean predicted distance shorter than the negative pairs ($-0.03$ vs. $-1.46$ and $-0.01$ vs. $-1.04$). This indicates that the model learns to separate related and unrelated alert groups. 
In \autoref{fig:pred_closest}, we plot the alert groups predicted as the most similar to the ones in the knowledge base for \texttt{escalated\_sudo\_command} (left) and \texttt{attacker\_change\_user} (right). 
Here we see that the model is able to separate true positives with only a few false positives.

\begin{table}[h]
    \centering
    \begin{tabular}{|l|r|r|r|r|r|r|}
        \hline
             & \multicolumn{2}{c|}{\makecell{\textbf{Escalated} \\ \textbf{Sudo Command}}} & \multicolumn{2}{c|}{\makecell{\textbf{Attacker} \\ \textbf{Change User}}} & \multicolumn{2}{c|}{\textbf{WPscan}} \\
        \hline
             & \textbf{Positive} & \textbf{Negative} & \textbf{Positive} & \textbf{Negative} & \textbf{Positive} & \textbf{Negative} \\
        \hline
        Number of graphs & 2 & 36681 & 2 & 36681 & 2 & 100 \\
        \hline
        Mean distance & -0.03 & -1.46 & -0.01 & -1.04 & -1.73 & -82401 \\
        \hline
        Standard deviation & 0.02 & 13.81 & 0.00 & 10.11 & 0.58 & 81456 \\
        \hline
        Longest distance & -0.05 & -1103.30 & -0.01 & -917.97 & -2.31 & -270081 \\
        \hline
        Shortest distance & -0.00 & -0.00 & -0.00 & -0.07 & -1.15 & -1828 \\
        \hline
    \end{tabular}
    \caption{Predicted distance for positive and negative alert group pairs}
    \label{tab:GMN_results}
\end{table}

\vspace{-10pt}
Graphs from \texttt{wpscan} are much larger than the two other classes, and the resulting comparison is much slower. We therefore sample $100$ graphs for the negative examples. In the \texttt{wpscan} columns of \autoref{tab:GMN_results}, we observe that the model is still able to separate alert groups representing \texttt{wpscan} from unrelated alert groups. However, issues with scalability become more apparent as comparisons are slow. 
% Comparing 100 alert groups takes 11m 53s on a Intel(R) Xeon(R) CPU E5-2680 v4 @ 2.40GHz cpu
%\FloatBarrier

\section{Discussion}
\label{sec:discussion}
% alert group quality
Graph-based alert aggregation performs well in combining alerts related to the same attack step. Although it achieves a high cluster purity score, it cannot avoid all unrelated alerts. Achieving perfect purity requires knowing if an alert is a false positive, an exceedingly difficult challenge. 

% overall observations and size
The graph matching approach shows promise in relating alert groups for attack types \texttt{escalated\_sudo\_command} and \texttt{attacker\_change\_user}. These are attacks that result in medium-sized graphs with the number of nodes between $15$ and $100$. 
When relating alert groups of type \texttt{wpscan}, scalability challenges become apparent due to the number of nodes in \texttt{wpscan} alert groups ranging from $1,400$ to $13,300$.
% Comparison is slow
The computational cost of GMNs is high as a separate comparison is done for each pair of alert groups. This means that the number of comparisons needed is $I*K$ where $I$ is the number of incoming alert groups, and $K$ is the number of known alert groups. Additionally, the comparison of two graphs has the computational cost of $O(|V1||V2|)$  where $|V_1|$ and $|V_2|$ are the number of nodes in the two graphs, respectively~\cite{li2019graph}. 

Future work should focus on mitigating this. One approach is to use a simpler \emph{Graph Embedding Network} (GEN)~\cite{li2019graph}, as it allows all alert groups to be embedded in the same space, but this comes at the cost of comparison quality. A hybrid approach, using GENs for coarse filtering followed by GMN for fine-grained matching on a smaller candidate set, could balance speed and accuracy. Furthermore, actively managing the content of the knowledge base by identifying representative \emph{template graphs} for each attack step will reduce the number of comparisons. One interesting approach would be to search for a template graph using evolutionary algorithms~\cite{back1993overview}, where the GMN distance function can be used as the fitness metric.

% The knowledge base is important - enrichment and concept drift 
A central component of this approach to analysing alert groups is the knowledge base of previous alert groups. The capability of linking an alert groups to past alert groups is only as good as the quality of what you are linking it to. Managing this knowledge base is therefore important. As new attacks emerge, the knowledge base must be updated to incorporate them, while obsolete attacks should be removed or at least deprioritised. Actively managing this can help mitigate concept drift~\cite{webb2016characterizing}.
% enrichment
The knowledge base also serves as a resource for enrichment. By integrating relevant information into it prior to an incident, connections can be established instantly when it is most important.

\section{Conclusion} 
\label{sec:conclusion}
% one sentence
We have addressed the challenge of understanding and contextualising security alerts within an SOC, which is 
% Problem statement
a challenging task due to the large volume of alerts requiring rapid examination, correlation, and understanding.

% key findings 
We have presented an approach to correlate alerts in a structured way as \emph{graph-based alert groups} based on \emph{pivoting} around alert properties with the goal of capturing higher-level behavioural patterns. This data structure lays the foundation for a more in-depth analysis of incidents.  

Building on this, we have employed a graph-based machine learning approach to correlate new graph-based alert groups with known ones (stored in a knowledge base). This enables analysts to leverage historical context to gain additional information. This knowledge base provides a good way of maintaining knowledge that can automatically be correlated with new alert groups.

Through experimentation, we have demonstrated that by carefully selecting timeline-defining properties and time cut-offs, it is possible to combine related alerts into a cohesive graph-based alert group corresponding to a specific attack step. 
The evaluation showed that our graph-based approach outperformed a simpler time-based approach in terms of group purity and recognisability. These experiments have provided promising results for our first (RQ1) and second (RQ2) research questions.

To demonstrate the usability for downstream machine learning tasks, we employed a Graph Matching Network (GMN) to correlate new alert groups with a knowledge base of historical incidents. Our experiments demonstrated that GMNs can successfully identify similarities between current and historical alert groups for attack types resulting in medium-sized graphs ($10-100$ nodes). 
However, performance in terms of comparison speed diminished with larger graphs. We have therefore provided positive results for our third research question (RQ3) for medium-sized graphs, with some limitations identified for large graphs.

% % Further works
% Promising directions for future work are exploring other approaches to utilising the representation of graph-based alert groups and addressing the computational cost of Graph Matching Networks. 
% by creating an ideal \emph{template graph} for each of the attack steps (to use in the knowledge base).

To conclude, the two main contributions of this paper are: (1) a formal approach for constructing meaningful graph-based alert groups from raw alert streams; and (2) a demonstration of the capability of Graph Matching Networks for correlating these groups with historical incidents. This enables more abstract, behaviour-oriented analysis and automated contextualisation of alerts.

\paragraph{Acknowledgements} This work was partially funded by the European Union as part of the European Defence Fund (EDF) project AInception (GA No. 101103385). Views and opinions expressed are, however, those of the authors only and do not necessarily reflect those of the European Union (EU). The EU cannot be held responsible for them.

This preprint has not undergone peer review or any post-submission improvements or corrections. The Version of Record of this contribution will be published in Information Security: 28th International Conference, ISC 2025, Seoul, South Korea. The Springer version can be accessed at https://link.springer.com/book/978-3-032-08123-0 upon publication.

\bibliographystyle{splncs04}
\bibliography{refs}

\end{document}